\documentclass[aps,pra,twocolumn,showpacs,amsmath]{revtex4}
\usepackage{bm}

\begin{document}
\newcommand{\Tr}{\mathop{\mathrm{Tr}}\nolimits}
\font\viiirm=cmr8
\font\vector=cmmib10 
\def\vec#1{\hbox{\vector #1}}
\def\vecomega{\hbox{\vector\char"21}}
\def\vectheta{\hbox{\vector\char"12}}
\def\vecsigma{\hbox{\vector\char"1B}}
\def\vectau{\hbox{\vector\char"1C}}
\def\sXA{{\hbox{\xirm XA}}}
\def\sXB{{\hbox{\xirm XB}}}
\def\sA{{\hbox{\xirm A}}}\def\sB{{\hbox{\xirm B}}}\def\sX{{\hbox{\xirm X}}}
\font\xrm=cmr10
\def\ssXA{{\hbox{\xrm XA}}}
\def\ssXB{{\hbox{\xrm XB}}}
\def\ssA{{\hbox{\xrm A}}}\def\ssB{{\hbox{\xrm B}}}\def\ssX{{\hbox{\xrm X}}}
\def\upX{\up_\sX}\def\downX{\down_\sX}
\def\upA{\up_\sA}\def\downA{\down_\sA}
\def\upB{\up_\sB}\def\downB{\down_\sB}
\mathchardef\up="0222 \mathchardef\down="0223
\mathchardef\bra="0268 \mathchardef\ket="0269
\def\lastmodified{\the\year.\the\month.\the\day}

\title{A Purification Scheme and Entanglement Distillations}
\author{Hiromichi Nakazato}
\email{hiromici@waseda.jp}
\author{Makoto Unoki}
\author{Kazuya Yuasa}
\email{yuasa@hep.phys.waseda.ac.jp}
\affiliation{Department of Physics, Waseda University, Tokyo
169--8555, Japan}
\date[revised: ]{\lastmodified}
\begin{abstract}
A purification scheme which utilizes the action of repeated measurements
on a (part of a total) quantum system is briefly reviewed and is
applied to a few simple systems to show how it enables us to extract an
entangled state as a target pure state.
The scheme is rather simple (e.g., we need not prepare a
specific initial state) and is shown to have wide applicability and
flexibility, and is able to accomplish both the maximal fidelity and
non-vanishing yield.
\end{abstract}
\maketitle

\section{Introduction}\label{secI}
It is well known that in quantum mechanics, the action of measurement
affects the dynamics of the system just measured in an essential way.
This has to be contrasted with the situation in classical mechanics,
where the effect of measurement can be made as small as one wishes.
Typical phenomenon reflecting such a peculiarity in quantum mechanics
has been known under the name of ``Quantum Zeno Effect (QZE)"~\cite{MS77}
and has been extensively studied recently.
It states that if a system is frequently measured to confirm that it
is in its initial state, the change of the state based on the
Hamiltonian of the system is decelerated, or to state differently,
the decay of an unstable state is hindered by frequent measurements.
It is widely recognized, however, that there is no paradoxical point
in such phenomena and the effect is solely understood quantum
mechanically~\cite{dynQZE}.
Indeed, the projective measurement is not essential and is just
replaced with the generalized spectral decompositions~\cite{W63},
and the effect is due to the peculiar short-time dynamics of quantum
systems, known as the ``flat derivative" of the survival probability,
$\dot P(0)=0$.
The effect has also been examined experimentally and the first
announcement of its confirmation appeared in an oscillating
system~\cite{IHBW90} and then in a truly unstable system~\cite{FGR01}.
It is worth while mentioning that Raizen's group has observed the
deviation from the exponential law at short times~\cite{W97} and even
the acceleration of decay when the measurements are not frequent
enough~\cite{FGR01}.
The latter effect is called the ``Inverse (or Anti) Zeno Effect (IZE),"
which has also been studied and discussed~\cite{IZE}.
The QZE (and/or IZE) has still been explored extensively in the hope of
realizing a protection scheme for system's coherence against
possible decoherence by this (or its related) mechanism~\cite{FNPTT04}.

Here another interesting consequence of the peculiarity of quantum
measurement shall be disclosed.
We consider a total system that is composed of two (sub)systems A and B
and measurements shall be repeatedly performed only on system A at
regular intervals.
Our interest lies in the (asymptotic) dynamics of system B, which is
indirectly affected by the measurements on A through its interaction
with A\@.
It is shown that such indirect measurements can drive system B to a pure
state ({\it purification\/}), irrespectively of its initial state
that is mixed in general~\cite{NTY03}.
We can say that the effect of measurement is far reaching and profound:
It can control even the other parts of the system which are not touched
directly.
In the following sections, the mechanism of this purification
is briefly reviewed (Sec.~\ref{secII}) and is applied to a simple qubit
system (Sec.~\ref{secIII}) to show how it works and how it can be
optimized.
Since the entangled state, which is one of the key elements in
quantum technologies like quantum computation, information,
teleportation, etc.~\cite{qtec}, is a pure state of a compound system,
this method is used to extract entangled states~\cite{YNU04,NUY04}
in Secs.~\ref{secIV} and \ref{secV}.
A brief summary is presented in Sec.~\ref{secVI}.

\section{General Framework of Purification Through Repeated Measurements}
\label{secII}
Let the total system consist of two parts, system A and system B, and
the dynamics be described by the total Hamiltonian
\begin{equation}
H=H_A+H_B+H_{int},
\label{eq:HA+B}
\end{equation}
where $H_{int}$ stands for the interaction between the two (sub)systems.
We initially prepare the system in a product state
\begin{equation}
\rho_0=|\phi\rangle\langle\phi|\otimes\rho_B(0)
\label{eq:rho0}
\end{equation}
at $t=0$.
Such a state can be realized, say, if the system A is found in
the state $|\phi\rangle$ after the zeroth measurement.
Notice that system B can be in an {\it arbitrary mixed\/} state $\rho_B(0)$.
We perform measurements on A at regular intervals $\tau$ to confirm that it
is still in the state $|\phi\rangle$, even though the total system A+B
evolves unitarily in terms of the total Hamiltonian $H$.
Since the measurement is performed only on system A, the action
of such a (projective, for simplicity) measurement can be conveniently
described by the following projection operator
\begin{equation}
{\cal O}\equiv|\phi\rangle\langle\phi|\otimes\hat{1}_B.
\label{eq:calO}
\end{equation}
Thus the state of system A is set back to $|\phi\rangle$ every after
$\tau$, while that of B just evolves dynamically on the basis of the
total Hamiltonian $H$.
We repeat the same measurement, represented by (\ref{eq:calO}), $N$ times
and collect only those events in which system A has been found in
state $|\phi\rangle$ consecutively $N$ times;
other events are discarded.
The state of system B is then described by the density matrix
\begin{equation}
\rho_B^{(\tau)}(N)
=\bigl(V_\phi(\tau)\bigr)^N\rho_B(0)
 \bigl(V_\phi^\dagger(\tau)\bigr)^N/P^{(\tau)}(N),
\label{eq:rhoBN}
\end{equation}
where
\begin{equation}
V_\phi(\tau)\equiv\langle\phi|e^{-iH\tau}|\phi\rangle
\label{eq:Vphi}
\end{equation}
is an operator acting on B and
\begin{eqnarray}
P^{(\tau)}(N)
&\!=\!&{\rm Tr}\Bigl[({\cal O}e^{-iH\tau}{\cal O})^N\rho_0
 ({\cal O}e^{iH\tau}{\cal O})^N\Bigr]\nonumber\\
&\!=\!&{\rm Tr}_B\Bigl[\bigl(V_\phi(\tau)\bigr)^N\rho_B(0)
  \bigl(V_\phi^\dagger(\tau)\bigr)^N\Bigr]
\label{eq:PtauN}
\end{eqnarray}
is the success probability for these events to occur (yield).
This normalization factor in (\ref{eq:rhoBN}) reflects the fact that
only right outcomes are collected in this process.

In order to examine the asymptotic state of system B,
consider the spectral decomposition of the
operator $V_\phi(\tau)$, which is not hermitian,
$V_\phi(\tau)\not=V_\phi^\dagger(\tau)$.
We therefore need to set up both the right- and left-eigenvalue problems
\begin{equation}
V_\phi(\tau)|u_n)=\lambda_n|u_n),\qquad
(v_n|V_\phi(\tau)=\lambda_n(v_n|.
\label{eq:unvn}
\end{equation}
The eigenvalue $\lambda_n$ is complex valued in general, but its
absolute value is bounded~\cite{NUY04}
\begin{equation}
\quad0\leq|\lambda_n|\leq1,
\label{eq:bnded}
\end{equation}
which is a reflection of the unitarity of the time evolution operator
$e^{-iH\tau}$\@.
These eigenvectors are assumed to form a complete orthonormal set in
the following sense
\begin{equation}
\sum_n|u_n)(v_n|=\hat1_B,\qquad(v_n|u_m)=\delta_{nm}.
\label{eq:conset}
\end{equation}
Then the operator $V_\phi(\tau)$ itself is expanded in terms of these
eigenvectors
\begin{equation}
V_\phi(\tau)=\sum_n\lambda_n|u_n)(v_n|.
\label{eq:Vphiuv}
\end{equation}
It is now easy to see that the $N$th power of this operator is
expressed as
\begin{equation}
\bigl(V_\phi(\tau)\bigr)^N=\sum_n\lambda_n^N|u_n)(v_n|
\label{eq:VphiN}
\end{equation}
and therefore it is dominated by a single term for large $N$
\begin{equation}
\bigl(V_\phi(\tau)\bigr)^N
\xrightarrow{\text{large }N}
\lambda_0^N|u_0)(v_0|,
\label{eq:Vinf}
\end{equation}
when the largest (in magnitude) eigenvalue $\lambda_0$ is
{\it discrete, nondegenerate and unique\/}.
If these conditions are satisfied, the density operator of system B
is driven to a pure state
\begin{equation}
\rho_B^{(\tau)}(N)
\xrightarrow{\text{large }N}
|u_0)(u_0|/(u_0|u_0)
\label{eq:rhoBinf}
\end{equation}
with the probability
\begin{equation}
P^{(\tau)}(N)
\xrightarrow{\text{large }N}
|\lambda_0|^{2N}(u_0|u_0)(v_0|\rho_B(0)|v_0).
\label{eq:Pinf}
\end{equation}
The pure state $|u_0)$, which is nothing but the right-eigenvector of
the operator $V_\phi(\tau)$ belonging to the largest (in magnitude)
eigenvalue $\lambda_0$, is thus distilled in system B\@.
This is the purification scheme proposed in \cite{NTY03}.

A few comments are in order.
First, the final pure state $|u_0)$ toward which system B is to
be driven is dependent on the choice of the state $|\phi\rangle$ on
which system A is projected every after measurement, the
measurement interval $\tau$ and the Hamiltonian $H$, but does not
depend on the initial state of system B at all.
In this sense, the purification is accomplished irrespectively of the
initial (mixed) state $\rho_B(0)$.
Second, as is clear in the above exposition, what is crucial in this
purification scheme is the repetition of one and the same measurement
(more appropriately, spectral decomposition) and the measurement
interval $\tau$ need not be very small.  It remains to be an
adjustable parameter.
Third, if we can make other eigenvalues than $\lambda_0$ much smaller
in magnitude
\begin{equation}
|\lambda_n/\lambda_0|\ll1\quad\hbox{for } n\not=0,
\label{eq:quick}
\end{equation}
by adjusting parameters, we will need fewer steps (i.e., smaller $N$)
to purify system B\@.

It is now evident that the purification can be made {\it optimal\/},
if the conditions (\ref{eq:quick}) and
\begin{equation}
|\lambda_0|=1
\label{eq:noloss}
\end{equation}
are satisfied.
This condition (\ref{eq:noloss}) assures that we can repeat as many
measurements as we wish without running the risk of losing the
yield (success probability) $P^{(\tau)}(N)$ in order to make the
fidelity to the target state $|u_0)$,
\begin{equation}
F^{(\tau)}(N)\equiv{\rm Tr}_B
\left[\rho_B^{(\tau)}(N)|u_0)(u_0|/(u_0|u_0)\right],
\label{eq:F}
\end{equation}
higher.
Actually, the yield $P^{(\tau)}(N)$ decays like
\begin{align}
P^{(\tau)}(N)
&=\sum_{n,m}\lambda_n^N\lambda_m^*{}^N(v_n|\rho_B(0)|v_m)(u_m|u_n)
\nonumber\\
&
\xrightarrow{\text{large }N}
|\lambda_0|^{2N}(u_0|u_0)(v_0|\rho_B(0)|v_0)
\label{eq:Pasymp}
\end{align}
and the condition (\ref{eq:noloss}) can bring us with the non-vanishing
yield $(u_0|u_0)(v_0|\rho_B(0)|v_0)$ even in the $N\to\infty$ limit.
Therefore the condition (\ref{eq:noloss}) makes the two (sometimes not
compatible) demands, i.e., {\it higher fidelity and non-vanishing
yield\/}, achievable, with fewer steps when the condition
(\ref{eq:quick}) is met.
In this sense, the purification is considered to be optimal.

It would be desirable if an optimal purification can be realized by
an appropriate choice of the state $|\phi\rangle$ and/or tuning of
the measurement interval $\tau$ and parameters in a given system.
In the following sections, a few simple systems are examined to show
how such optimal purifications are made possible.

\section{Purification of a Qubit}\label{secIII}
As a simplest example, let us consider a total system of interacting
two qubits.
Systems A and B are represented by two qubits A and B, respectively,
and their two levels can be described as the two degrees of freedom
of spin-1/2 particle.
We measure qubit A at regular intervals $\tau$ and examine the state
of qubit B, which is in interaction with A\@.
The measurement is conveniently parameterized as that of a spin-1/2
particle along a particular direction $\vec n$.
After qubit A has been confirmed that its ``spin" is up along $\vec n$,
its state is projected to the eigenstate of the spin operator along
$\vec n$, i.e., $|\phi\rangle=\vec n\cdot\vectau|\phi\rangle$, with
$\vectau\equiv(\tau_1,\tau_2,\tau_3)$ being the Pauli matrices
acting on qubit A\@.
The state $|\phi\rangle$ is parameterized in terms of the two angles
$\theta$ and $\varphi$ as
\begin{equation}
|\phi\rangle=\cos(\theta/2)e^{-i{\varphi/2}}|\up\rangle
             +\sin(\theta/2)e^{i{\varphi/2}}|\down\rangle.
\label{eq:phi-n}
\end{equation}
It is an elementary task to write down the projected operator
$V_\phi(\tau)\equiv\langle\phi|e^{-iH\tau}|\phi\rangle$ in the
following form
\begin{equation}
V_\phi(\tau)=c_0+\vec c\cdot\vecsigma,
\label{eq:Vctau}
\end{equation}
where $\vecsigma\equiv(\sigma_1,\sigma_2,\sigma_3)$ are Pauli matrices
acting on qubit B and parameters $c_0$ and $\vec c\equiv(c_1,c_2,c_3)$
are complex valued.
Its eigenvalues $\lambda_\pm$ and the corresponding right- and
left-eigenvectors are easily found
\begin{equation}
\lambda_\pm=c_0\pm c,\qquad c\equiv\sqrt{{\vec c}^2},
\quad c_\pm\equiv c_1\pm ic_2,
\label{eq:lambdapm}
\end{equation}
\begin{eqnarray}
&&|u_+)={1\over\sqrt{2c(c-c_3)}}
        \Bigl[c_-|\up)+(c-c_3)|\down)\Bigr],\nonumber\\
&&|u_-)={1\over\sqrt{2c(c-c_3)}}
        \Bigl[(c_3-c)|\up)+c_+|\down)\Bigr],\\
\label{eq:upm}
&&(v_+|={1\over\sqrt{2c(c-c_3)}}
        \Bigl[c_+(\up|+(c-c_3)(\down|\Bigr],\nonumber\\
&&(v_-|={1\over\sqrt{2c(c-c_3)}}
        \Bigl[(c_3-c)(\up|+c_-(\down|\Bigr].
\label{eq:vpm}
\end{eqnarray}

Now let the total Hamiltonian of this system be given by
\begin{eqnarray}
&&H={\omega_A\over2}(1+\tau_3)+{\omega_B\over2}(1+\sigma_3)\nonumber\\
&&\phantom{H={\omega_A\over2}}
+g(\tau_+\sigma_-+{\rm h.c.})+h(\tau_+\sigma_++{\rm h.c.}),
\label{eq:H2q}
\end{eqnarray}
where real parameters $g$ and $h$ are responsible for the interaction
between the two qubits, A and B\@.
In this case, the parameters $c_0,\ldots,c_3$ in
(\ref{eq:Vctau})--(\ref{eq:vpm}) are explicitly calculated to be
\begin{align}
c_0&={1\over2}\left(\cos{\tau\theta_h\over2}+\cos{\tau\theta_g\over2}\right)
\nonumber\\
&\phantom{={1\over2}}
   -{i\over2}\left({\omega_+\over\theta_h}\sin{\tau\theta_h\over2}
    +{\omega_-\over\theta_g}\sin{\tau\theta_g\over2}\right)\cos\theta,
\label{eq:c0}\\
c_1&=-i\left({h\over\theta_h}\sin{\tau\theta_h\over2}
     +{g\over\theta_g}\sin{\tau\theta_g\over2}\right)\sin\theta\cos\varphi,
\label{eq:c1}\\
c_2&=i\left({h\over\theta_h}\sin{\tau\theta_h\over2}
     -{g\over\theta_g}\sin{\tau\theta_g\over2}\right)\sin\theta\sin\varphi,
\label{eq:c2}\\
c_3&={1\over2}\left(\cos{\tau\theta_h\over2}-\cos{\tau\theta_g\over2}\right)
     \cos\theta\nonumber\\
&\phantom{={1\over2}\left(\cos\right.}
   -{i\over2}\left({\omega_+\over\theta_h}\sin{\tau\theta_h\over2}
    -{\omega_-\over\theta_g}\sin{\tau\theta_g\over2}\right),
\label{eq:c3}
\end{align}
where we have introduced
\begin{equation}
\omega_\pm\equiv\omega_A\pm\omega_B,\quad
\theta_h\equiv\sqrt{\omega_+^2+4h^2},
\quad\theta_g\equiv\sqrt{\omega_-^2+4g^2}.
\label{eq:parameters}
\end{equation}

In order to illustrate how an optimal purification can be achieved
in this system, consider the case where we measure qubit A along the
3-direction, that is, we choose $\theta=0$ and $|\phi\rangle=|\up
\rangle$.
Then the eigenvalues $\lambda_\pm$ in (\ref{eq:lambdapm}) and the
corresponding eigenvectors are
\begin{eqnarray}
\lambda_+
=\cos{\tau\theta_h\over2}-i{\omega_+\over\theta_h}\sin{\tau\theta_h\over2}
&\Longleftrightarrow&|u_+)=|\up),\\
\label{eq:plus}
\lambda_-
=\cos{\tau\theta_g\over2}-i{\omega_-\over\theta_g}\sin{\tau\theta_g\over2}
&\Longleftrightarrow&|u_-)=|\down).
\label{eq:minus}
\end{eqnarray}
Since we have
\begin{equation}
|\lambda_+|^2=1-{4h^2\over\theta_h^2}\sin^2{\tau\theta_h\over2},
\quad
|\lambda_-|^2=1-{4g^2\over\theta_g^2}\sin^2{\tau\theta_g\over2},
\label{eq:abs}
\end{equation}
the purification can be made optimal, e.g., when the parameters are
adjusted so that $h\sin(\tau\theta_h/2)=0$ is satisfied.
In this case, $|\lambda_+|=1$ and qubit B is driven to a pure state
$|\up)$, more quickly for larger $g$ satisfying
$\sin^2(\tau\theta_g/2)=1$.
A similar situation can happen;
we can extract $|\down)$ in qubit B, more quickly for larger $h$
satisfying $\sin^2(\tau\theta_h/2)=1$, if we adjust parameters so that
$g\sin(\tau\theta_g/2)=0$ holds.

Needless to say, there are cases where such purifications are
not possible.
For example, consider a case where we measure qubit A in the 1-2 plane,
i.e., $\theta=\pi/2$.
In this case, since the parameter $c_0$ is real, while all the other
parameters $c_1,\,c_2$ and $c_3$ become pure imaginary, the eigenvalues
$\lambda_\pm=c_0\pm c$ are degenerated in magnitude
\begin{equation}
|\lambda_+|=\sqrt{c_0^2+|c|^2}=|\lambda_-|
\label{eq:degenerate}
\end{equation}
and no purification can occur in this particular case.

\section{Entanglement Distillation: I}\label{secIV}
As is mentioned in the Introduction, since entanglement is one of the
key elements in quantum technologies~\cite{qtec}, it would be useful
if the present scheme of purification can be used to extract an
entangled state as a target pure state.
Notice that since the target system has never been measured directly
in the present scheme, it is considered to be suited for
extraction of such a fragile pure state as an entangled state.
Actually any measurement on its subsystem that consits of
entanglement would result in the destruction of the entanglement.

In order to see an entanglement distillation on the basis of the
present idea of purification~\cite{YNU04,NUY04}, we consider a total
system composed of a compound system A+B, in which an entangled state
is to be extracted, and another system C\@.
Systems A and B interact with system C separately, but do not
interact directly with each other.
We measure system C repeatedly at regular intervals $\tau$ and endeavor
to extract an entangled state as a pure state in A+B\@.
In this section, a simple model, in which all systems A, B and C are
represented by qubits, is considered with a model Hamiltonian
\begin{eqnarray}
&&H={\Omega\over2}(1+\sigma_3^A)+{\Omega\over2}(1+\sigma_3^B)
  +{\omega\over2}(1+\tau_3)\nonumber\\
&&\phantom{H={\Omega\over2}}
    +g(\sigma_+^A\tau_-+\sigma_+^B\tau_-+{\rm h.c.})\nonumber\\
&&\phantom{H={\Omega\over2}+}
    +h(\sigma_+^A\tau_++\sigma_+^B\tau_++{\rm h.c.}),
\label{eq:HABC}
\end{eqnarray}
where the Pauli matrices $\tau_i$ act on system C\@.
It is assumed here for simplicity that the two systems A and B are the
same and the Hamiltonian is symmetric under the exchange
A$\leftrightarrow$B\@.

In order to find the spectral decomposition (\ref{eq:Vphiuv}) of the
projected operator $V_\phi(\tau)$ in this case, it turns out to be
convenient to introduce the Bell states
\begin{equation}
|\Psi^\pm)={1\over\sqrt2}
           \Bigl[|\up\down)\pm|\down\up)\Bigr],\qquad
|\Phi^\pm)={1\over\sqrt2}
           \Bigl[|\up\up)\pm|\down\down)\Bigr]
\label{eq:Bellstates}
\end{equation}
as a complete orthonormal set for A+B, because we are interested in
extraction of such entangled states.
Indeed, the eigenstates of the total Hamiltonian $H$ can be found
after their classification according to the above mentioned
A$\leftrightarrow$B symmetry and a ``parity" ${\cal P}\equiv
\sigma_3^A\sigma_3^B\tau_3$:\par
\noindent
1) A$\leftrightarrow$B symmetric and ${\cal P}=+$
\begin{equation}
H\begin{pmatrix}|\Phi^+\up\rangle\\
          |\Phi^-\up\rangle\\
          |\Psi^+\down\rangle\end{pmatrix}
=\begin{pmatrix}\Omega+\omega&\Omega&g+h\\
          \Omega&\Omega+\omega&-g+h\\
          g+h&-g+h&\Omega\end{pmatrix}
 \begin{pmatrix}|\Phi^+\up\rangle\\
          |\Phi^-\up\rangle\\
          |\Psi^+\down\rangle\end{pmatrix},
\label{eq:S+}
\end{equation}
\noindent
2) A$\leftrightarrow$B symmetric and ${\cal P}=-$
\begin{equation}
H\begin{pmatrix}|\Phi^+\down\rangle\\
          |\Phi^-\down\rangle\\
          |\Psi^+\up\rangle\end{pmatrix}
=\begin{pmatrix}\Omega&\Omega&g+h\\
          \Omega&\Omega&g-h\\
          g+h&g-h&\Omega+\omega\end{pmatrix}
 \begin{pmatrix}|\Phi^+\down\rangle\\
          |\Phi^-\down\rangle\\
          |\Psi^+\up\rangle\end{pmatrix},
\label{eq:S-}
\end{equation}
\noindent
3) A$\leftrightarrow$B anti-symmetric and ${\cal P}=-$
\begin{equation}
H|\Psi^-\up\rangle=(\Omega+\omega)|\Psi^-\up\rangle,
\label{eq:A-}
\end{equation}
\noindent
4) A$\leftrightarrow$B anti-symmetric and ${\cal P}=+$
\begin{equation}
H|\Psi^-\down\rangle=\omega|\Psi^-\down\rangle.
\label{eq:A+}
\end{equation}
Here $|\Phi^+\up\rangle\equiv|\Phi^+)\otimes|\up\rangle$, etc.
Thus the time evolution operator $e^{-iH\tau}$ is expressed as
\begin{align}
e^{-iH\tau}=&\sum_se^{-iE_s\tau}|s\rangle\langle s|\nonumber\\
  &+|\Psi^-)(\Psi^-|
  \Bigl[e^{-i(\Omega+\omega)\tau}|\up\rangle\langle\up|
  +e^{-i\Omega\tau}|\down\rangle\langle\down|\Bigr],
\label{eq:eHtau}
\end{align}
where the summation is taken over the six A$\leftrightarrow$B
symmetric eigenstates of $H$, denoted as $|s\rangle$, that are given as
linear combinations of the six states in (\ref{eq:S+}) and (\ref{eq:S-}).
Owing to the A$\leftrightarrow$B symmetry of $H$, the A$\leftrightarrow$B
anti-symmetric state $|\Psi^-)$ does not mix with the other
(A$\leftrightarrow$B symmetric) eigenstates.

We are now in a position to examine the spectrum of the projected
operator $V_\phi(\tau)\equiv\langle\phi|e^{-iH\tau}|\phi\rangle$.
If the measurement of C projects its state on
\begin{equation}
|\phi\rangle=\alpha|\up\rangle+\beta|\down\rangle,
\label{eq:phiC}
\end{equation}
the operator reads
\begin{eqnarray}
&&V_\phi(\tau)=\langle\phi|e^{-iH\tau}|\phi\rangle\nonumber\\
&&\phantom{V_\phi(\tau)}
 =\sum_se^{-iE_s\tau}
 \Bigl[|\alpha|^2\langle\up|s\rangle\langle s|\up\rangle
  +|\beta|^2\langle\down|s\rangle\langle s|\down\rangle
  \nonumber\\
&&\phantom{V_\phi(\tau)=\sum_se^{-iE_s\tau}\left[|\alpha|^2\right.}
 +\alpha^*\beta\langle\up|s\rangle\langle s|\down\rangle
 +{\rm h.c.}\Bigr]\nonumber\\
&&\phantom{V_\phi(\tau)=\sum_s}
 +|\Psi^-)(\Psi^-|
  \left[|\alpha|^2e^{-i(\Omega+\omega)\tau}
  +|\beta|^2e^{-i\Omega\tau}\right].\nonumber\\
\label{eq:VphiAB}
\end{eqnarray}
From this expression, it is evident that the Bell state $|\Psi^-)$ is
always one of the eigenstates of this operator and if the
measurement interval $\tau$ is so adjusted that the condition
$\omega\tau=2\pi$ is met, its eigenvalue $\lambda_{\Psi^-}$ becomes
maximum in magnitude
\begin{equation}
\omega\tau=2\pi\quad\longrightarrow\quad|\lambda_{\Psi^-}|=1,
\label{eq:noloss1}
\end{equation}
irrespectively of the projected state $|\phi\rangle$ of system C
because $|\alpha|^2+|\beta|^2=1$.
This clearly demonstrates a possibility of entanglement distillation
in this simple system, in the sense that one of the conditions for
optimal purification (\ref{eq:noloss}) can be realized by
(\ref{eq:noloss1}) and the entangled state $|\Psi^-)$ would surely be
extracted, only if the other eigenvalues of the operator than
$\lambda_{\Psi^-}$ can be made (much) smaller in magnitude.
Further details of the analysis of such conditions for the entanglement
distillation and its optimization are found in \cite{YNU04} and
\cite{NUY04} in a slightly simplified case.

\section{Entanglement Distillation: II}\label{secV}
The example in the previous section explicitly demonstrates that we
can distill an entangled state in the system A+B, through the repeated
measurements on the other system C that separately interacts with A and
B\@.
The framework is rather simple and the distillation can be made
optimal.
There is, however, a kind of drawback in this scheme.
As is clear in its exposition, it is assumed that system C, on which
the measurement is performed, {\it always and simultaneously
interacts\/} with both A and B and these interactions 
are crucial for the entanglement distillation.
Stating differently, systems A and B (and C) are not (and/or will not
be) able to be {\it separated spatially\/}, which implies that no
entanglement between spatially separated systems is possible by the
scheme presented in Sec.~\ref{secIV}\@.
It would not be suited to the situations where entanglements
among spatially separated systems are required, as in quantum
teleportation.

In this section, a resolution to this problem is presented.
Since the two systems, A and B, an entanglement between which is to
be driven, are considered to be placed at different places, let us
consider, instead of system C which can no longer interact
simultaneously with A and B, another quantum system, say X, which is
assumed to interact with A and B, not simultaneously, but {\it
successively\/}~\cite{M02}.
System X plays the role of an ``entanglement mediator."
After such successive interactions with A and then B, system X is
measured to confirm that it is in a certain state.
If system X is found in this particular state, X is again brought to
interaction with A and then with B.
This process, i.e., X's interaction with A, that with B and
measurement on X, will be repeated many ($N$) times and we are
interested in the asymptotic state of system A+B in the hope of
distilling an entangled state.

There are a couple of points to be mentioned here.
First, it is clear that in spite of these modifications, the
new scheme presented here shares essentially the same idea of
purification with the previous ones:
The dynamics of the system can be affected, in an essential way, by
the action of measurement, even if its effect is not direct.
Second, such a successive interaction would be conveniently treated
in terms of a time-dependent (effective) Hamiltonian $H(t)$.
We may thus avoid possible complications caused by the introduction
of spatial degrees of freedom, still keeping the essential points.

In order to see how the new scheme works, consider again a
three-qubit system, A+B+X, for definiteness and simplicity.
We prepare system X, say in up state \hbox{$|\up\rangle$}, while
the system A+B can be in an arbitrary mixed state.
It is assumed that systems A and B are spatially separated and
have no contact with each other and that only system X can interact
with them locally for definite time durations.
Now consider the following process:
\begin{enumerate}
\item System X is first brought to interaction with system A for time
duration $t_A$.
The Hamiltonian here is given by $H(t)=H_0+H_{XA}$.
Then the interaction is switched off and the total system evolves
freely with the free Hamiltonian $H_0$ for $\tau_A$.
\item System X then interacts with system B for time duration $t_B$,
the dynamics of which is now described by another Hamiltonian
$H(t)=H_0+H_{XB}$.
After that, the total system again evolves freely with the Hamiltonian
$H_0$ for $\tau_B$.
\item A (projective) measurement is performed on system X to select
only up state $|\up\rangle$.
Other states are discarded.
\end{enumerate}
Then this process is repeated $N$ times: 1$\to$2$\to$3$\to$1$\to\cdots\to$1$\to$2$\to$3.

It is shown below that the following choice of the Hamiltonians
\begin{eqnarray}
&\displaystyle
 H_0={\omega\over2}(1+\sigma_3^A)+{\omega\over2}(1+\sigma_3^B)
    +{\omega\over2}(1+\sigma_3^X),&\nonumber\\
&H_{XA}=g_A\sigma_1^X\sigma_1^A,\qquad
H_{XB}=g_B\sigma_1^X\sigma_1^B&
\label{eq:HABX}
\end{eqnarray}
actually results in an entanglement distillation in system A+B\@.
It is important to notice that the above choice of the interaction
Hamiltonians is closely connected to the details of the process
1$\to$2$\to$3 and another choice, e.g., $\sigma_+^X\sigma_-^{A(B)}$
for $H_{XA(B)}$, would result not in an entangled state, but in a
product state, in this particular process.

The next task is to find the spectral decomposition of the
projected operator $V_\up$ defined, in this case, by
\begin{eqnarray}
&\displaystyle
V_\up=\langle\up|e^{-iH_0\tau_B}e^{-i(H_0+H_{XB})t_B}
\phantom{e^{-i(H_0+H_{XA})t_A}|\up\rangle}&\nonumber\\
&\phantom{V_\up=\langle\up|e^{-iH_0\tau_B}}
\displaystyle
\times e^{-iH_0\tau_A}e^{-i(H_0+H_{XA})t_A}|\up\rangle.&
\label{eq:Vup}
\end{eqnarray}
Since a parity defined by ${\cal P}\equiv\sigma_3^A\sigma_3^B$ is
conserved in this system, eigenstates of the operator $V_\up$
are easily found.
Indeed, we can classify every state of system A+B into two sectors
according to the parity $\cal P=\pm$ and the action of the operator
$V_\up$ is closed within each sector.
For $\cal P=+$ states, the action is represented by a matrix $\cal M$
\begin{equation}
V_\up\left[\begin{matrix}|\up\up)\\
                              |\down\down)\end{matrix}\right]
=e^{-i\omega(t_A+\tau_A+t_B+\tau_B)}{\cal M}
\left[\begin{matrix}|\up\up)\\
                    |\down\down)\end{matrix}\right],
\label{eq:P+}
\end{equation}
where its matrix elements read
\begin{eqnarray}
&&{\cal M}_{11}
=e^{-i\omega(t_A+2\tau_A+t_B+2\tau_B)}
 (\cos\zeta_A-i\sin\zeta_A\cos2\xi_A)\nonumber\\
&&\phantom{{\cal M}_{11}=e^{-i\omega(t_A)}}
\times(\cos\zeta_B-i\sin\zeta_B\cos2\xi_B),\nonumber\\
&&{\cal M}_{12}
=-e^{-i\omega t_A}\sin\zeta_A\sin2\xi_A\sin g_B t_B,\nonumber\\
&&{\cal M}_{21}
=-e^{-i\omega(t_B+2\tau_B)}
 \sin g_A t_A \sin\zeta_B\sin2\xi_B,\nonumber\\
&&{\cal M}_{22}
=\cos g_A t_A\cos g_B t_B,
\label{eq:Mij}
\end{eqnarray}
while, for $\cal P=-$ states, it is represented by another matrix
$\cal N$
\begin{equation}
V_\up\left[\begin{matrix}|\up\down)\\
                              |\down\up)\end{matrix}\right]
=e^{-i\omega(t_A+t_B+2\tau_B)}{\cal N}
\left[\begin{matrix}|\up\down)\\
                    |\down\up)\end{matrix}\right],
\label{eq:P-}
\end{equation}
with its matrix elements
\begin{eqnarray}
&&{\cal N}_{11}
=e^{-i\omega(2\tau_A+t_B)}
 (\cos\zeta_A-i\sin\zeta_A\cos2\xi_A)\cos g_B t_B,\nonumber\\
&&{\cal N}_{12}
=-\sin\zeta_A\sin2\xi_A\sin\zeta_B\sin2\xi_B,\nonumber\\
&&{\cal N}_{21}
=-e^{-i\omega(t_A+2\tau_A+t_B)}\sin g_A t_A\sin g_B t_B,\nonumber\\
&&{\cal N}_{22}
=e^{-i\omega(t_A+2\tau_A)}
 \cos g_A t_A(\cos\zeta_B-i\sin\zeta_B\cos2\xi_B).\nonumber\\
\label{eq:Nij}
\end{eqnarray}
Here the angles are defined by
\begin{equation}
\zeta_{A(B)}=t_{A(B)}\sqrt{\omega^2+g_{A(B)}^2},\quad
\tan2\xi_{A(B)}={g_{A(B)}\over\omega}.
\label{eq:angles}
\end{equation}

In order to see the possibility of entanglement
distillation in this framework, it is enough to consider a much more
simplified case.
Let the two systems A and B be treated symmetrically, that is, all
parameters are taken to be the same for A and B
\begin{eqnarray}
&\displaystyle
g_A=g_B\equiv g,\quad t_A=t_B\equiv t,\quad\tau_A=\tau_B\equiv\tau,&
\nonumber\\
&\displaystyle
(\zeta_{A(B)}\to\zeta,\quad\xi_{A(B)}\to\xi).&
\label{eq:A=B}
\end{eqnarray}
It is then easy to see that if the parameters satisfy
\begin{equation}
\cos\zeta-i\sin\zeta\cos2\xi=-e^{i\omega\tau}\cos gt,
\label{eq:optimal}
\end{equation}
an optimal purification of an entangled state $|\Psi)$ of the form
\begin{equation}
|\Psi)\equiv{1\over\sqrt2}\Bigl[|\up\down)+e^{i\chi}|\down\up)\Bigr]
\label{eq:Psi}
\end{equation}
with $\chi=\omega(t+\tau)$ is actually possible, provided
\begin{equation}
\cos gt\sin gt\not=0,\quad
\omega(t+\tau)\not=2n\pi\quad(n\hbox{: integer}).
\label{eq:constraint}
\end{equation}
In fact, one can show that $|\Psi)$ is an eigenstate of the operator
$V_\up$
\begin{equation}
V_\up|\Psi)=\lambda_\Psi|\Psi).
\label{eq:VPsi}
\end{equation}
The eigenvalue $\lambda_\Psi$ is maximum in magnitude
\begin{equation}
\lambda_\Psi=-e^{-3i\omega(t+\tau)},\quad|\lambda_\Psi|=1,
\label{eq:lambdaPsi}
\end{equation}
while all the other eigenvalues remain smaller than unity in magnitude,
under the conditions (\ref{eq:optimal}) and (\ref{eq:constraint}).
Therefore, we can repeat the process 1$\to$2$\to$3 as many times as is
required to achieve the desired (high) fidelity, without reducing the
yield.

This is an example of (optimal) entanglement distillations, where the
entanglement between two qubit systems that are (or can be) spatially
separated, is extracted through their successive interactions with
another qubit, on which one and the same measurement is repeated
regularly.
Further details of this model and applications to other quantum
systems, e.g., extraction of entanglement between two cavity modes at
a distance, will be reported elsewhere.

\section{Summary}\label{secVI}
In this paper, a new purification scheme recently
proposed~\cite{NTY03} is applied to a few simple qubit systems to
explicity show its ability of qubit purification (Sec.~\ref{secIII})
and entanglement distillations (Secs.~\ref{secIV} and \ref{secV}) for
two-qubit systems.
The important and essential idea, on which these particular examples
are based, is to utilize the effect caused by the action of measurement
on quantum systems.

It should be stressed again that since the basic idea is so simple,
that is, one has only to repeat one and the same measurement without
being concerned about the preparation of a specific initial (pure) state,
this purification scheme is considered to have wide applicability and
flexibility.
Furthermore, it enables us to make the two demands---{\it the maximal
fidelity and non-vanishing yield\/}---compatible.
The examples presented in this paper just show these characteristics and
many variants can be devised according to the actual setups.

\begin{acknowledgments}
The authors acknowledge useful and helpful discussions with
Prof.~I. Ohba.
Fruitful discussions with P. Facchi and S. Pascazio are also
appreciated.
One of the authors (H.N.) is grateful for the warm hospitality at
Universit\`a di Palermo, where he enjoyed the inspiring discussions
with A. Messina's group.
This work is partly supported by a Grant for The 21st Century COE
Program (Physics of Self-Organization Systems) at Waseda University
and a Grant-in-Aid for Priority Areas Research (B) from the Ministry
of Education, Culture, Sports, Science and Technology, Japan
(No.~13135221), by a Grant-in-Aid for Scientific Research (C)
(No.~14540280) from the Japan Society for the Promotion of Science,
by a Waseda University Grant for Special Research Projects
(No.~2002A-567) and by the bilateral Italian-Japanese project 15C1 on
``Quantum Information and Computation'' of the Italian Ministry for
Foreign Affairs.
\end{acknowledgments}


\end{document}